\documentclass[aps,prb,superscriptaddress,twocolumn]{revtex4}
\usepackage{graphicx}
\usepackage{color}
\begin{document}

\title{Ferro-quadrupolar ordering in PrTi$_2$Al$_{20}$}

\author{Taku J Sato}
\author{Soshi Ibuka}
\author{Yusuke Nambu}
\author{Teruo Yamazaki}
\affiliation{Neutron Science Laboratory, Institute for Solid State Physics, University of Tokyo, 106-1 Shirakata, Tokai, Ibaraki 319-1106, Japan}
\author{Akito Sakai}
\author{Satoru Nakatsuji}
\affiliation{Institute for Solid State Physics, University of Tokyo, 5-1-5 Kashiwanoha, Chiba 277-8581, Japan}

\date{\today}

\begin{abstract}
Origin of the non-magnetic phase transition in PrTi$_2$Al$_{20}$, reported earlier in the macroscopic study, has been asserted microscopically using elastic and inelastic neutron scattering techniques.
It has been shown spectroscopically that the crystalline-electric-field ground state is a non-magnetic $\Gamma_3$ doublet, whereas the excited states are two triplets ($\Gamma_4$ and $\Gamma_5$) and a singlet ($\Gamma_1$).
The diffraction experiment under external magnetic field shows that the non-magnetic transition is indeed ferro-quadrupolar ordering, which takes place as a consequence of cooperative removal of the ground-state-doublet degeneracy.
It is therefore concluded that PrTi$_2$Al$_{20}$ is another rare example of Pr compounds exhibiting non-magnetic quadrupolar order.
\end{abstract}


\maketitle

\section{Introduction}
Even when the magnetic dipolar degree of freedom is completely suppressed by a crystalline-electric field (CEF) from surrounding atoms, some $4f$ elements can still have higher-rank-tensor degrees of freedom, such as electric quadrupoles, giving rise to mysterious low-temperature non-magnetic anomalies.~\cite{MorinP90}
Such anomalies include relatively simple ferro- and antiferro-quadrupolar ordering,~\cite{SuzukiH97,SuzukiO04,IwasaK02} as well as much elaborate incommensurately-modulated-quadrupolar order.~\cite{OnimaruT05}
In metals, those degrees of freedom may couple with conduction electrons.
In case of quadrupoles, the coupling results in a non-trivial quadrupolar Kondo effect, which is a very intriguing issue of active research.~\cite{CoxDL87,CoxDL98}
In reality, however, the quadrupolar (or higher-order) degree of freedom frequently coexists with dipolar terms, and thus its interplay with the dominant dipolar ordering results in much complicated behavior,~\cite{KuwaharaK07,TanakaY04,HirotaK00} prohibiting us to elucidate phenomena solely due to the higher-order degree of freedom.
Hence, $4f$ systems with a ground state possessing only higher-order degree of freedom are of particular interest.

Such a ground state may possibly be realized with a Pr$^{3+}$ ion in cubic CEF, since the non-magnetic, but quadrupolar- and octupolar-active, $\Gamma_3$ doublet may be the ground state for the $4f^2$ electronic configuration in a certain range of CEF parameters.~\cite{LeaKR62}
Therefore, many attempts have been made to find cubic Pr intermetallic compounds with the CEF ground state being the $\Gamma_3$ doublet.
A recent outcome of intensive material search is the ternary Pr$T_{2}X_{20}$ ($T =$ Ti, V, Ru, Ir, ..., and $X = $Zn and Al) intermetallic compounds.~\cite{OnimaruT10,SakaiA11,OnimaruT11}
Pr$T_2X_{20}$ belongs to the cubic space group $Fd\bar{3}m$, where the Pr$^{3+}$ ion, with the local site symmetry $\bar{4}3m$ ($T_d$), is located at the center of a perfect Friauf polyhedron with the coordination number 16, consisting of $X$ atoms.~\cite{NiemannS95,ThiedeVMT98}
A growing number of macroscopic studies on the Pr$T_2X_{20}$ compounds have revealed a variety of intriguing low-temperature phenomena in this system, ranging from possibly ferro- and antiferro-quadrupolar ordering to superconductivity.

In the present study we choose PrTi$_2$Al$_{20}$,~\cite{SakaiA11} one of the Pr$T_2X_{20}$ series compounds.
Magnetic susceptibility measurements on PrTi$_2$Al$_{20}$ show Curie-Weiss behavior at high temperatures $T > 250$~K, yielding an effective moment size of $\mu_{\rm eff} = 3.50\mu_{\rm B}$.
In the high-temperature range, the electrical resistivity shows increasing $\ln T$ behavior, a sign of magnetic Kondo effect, indicating a considerable coupling between the $4f$ and conduction electrons.
As temperature is decreased, the magnetic susceptibility becomes almost temperature independent below $T < 10$~K.
This suggests that the ground state is non-magnetic.
Nevertheless, the specific heat and the electric resistivity show clear anomaly at $T_{\rm c} \simeq 2$~K, and therefore, there should remain non-magnetic degree of freedom that orders at such a low temperature.
As the estimated entropy of $4f$ subsystem at $T = 5$~K reaches $R\ln 2$, it has been suggested that the ground state is the non-magnetic $\Gamma_3$ doublet.
Assuming this $\Gamma_3$ ground state, PrTi$_2$Al$_{20}$ could  have active quadrupolar or octupolar degree of freedom at sufficiently low temperatures, and hence one of those degrees of freedom may be the order parameter for the non-magnetic transition observed in the specific heat and resistivity measurements.
As the sharp specific-heat anomaly at $T_{\rm c}$ becomes strongly broadened under high magnetic field, ferro-quadrupolar ordering is inferred.~\cite{SakaiA11}

To confirm this scenario, the CEF splitting has to be determined using a spectroscopic technique.
However, up to now no spectroscopic study on the CEF splitting in PrTi$_2$Al$_{20}$ (nor even in any Pr$T_2X_{20}$ compounds) has been reported.
Furthermore, for conclusive understanding of the non-magnetic transition, it is essential to determine the symmetry of its order parameter, but this has not been explored experimentally at all.
In the present study, we, therefore, undertook neutron inelastic-scattering and diffraction experiments to determine the CEF splitting scheme, and to pin down the nature of the ordering degree of freedom in PrTi$_2$Al$_{20}$.

\section{Experimental details}
Single crystals of PrTi$_2$Al$_{20}$ were grown by an Al self-flux method under vacuum, using the pure starting elements, 99.99\%-Pr, 99.9\%-Ti, and 99.999\%-Al.
The largest single piece of grown crystals was approximately 50~mg, which was used in the single crystalline diffraction study.
As a polycrystalline sample, we have collected a few hundreds of tiny single crystals, amounting roughly 1~gram in total.

For the inelastic experiment, the polycrystalline sample was wrapped in an aluminum foil, and sealed in a standard aluminum sample can with He exchange gas.
The sample can was set to a $^{4}$He closed-cycle refrigerator, to a $^{3}$He closed-cycle refrigerator, or to a 6~T vertical field magnet, depending on necessity of the lowest temperature and the magnetic field.
The inelastic experiment has been carried out using the ISSP-GPTAS(4G) triple-axis spectrometer installed at JRR-3, Tokai, Japan.
Pyrolytic graphite (PG) 002 reflections were used for both the monochromator and analyzer, which were set in a horizontally-and-vertically (doubly) focusing condition to increase counting efficiency.
The collimations were 40'(Open)-3RC-7RC-30S, where 3RC, 7RC and 30S stand for radial collimators with 3 and 7 blades, and a slit with 30~mm opening before the detector.
The final neutron energy was fixed to 14.7~meV, and a PG filter was inserted between the 3rd collimator and analyzer to eliminate the higher harmonic neutrons.

For the diffraction experiment, the 50~mg single crystal was used with the 100 and 010 axes in the scattering plane.
The single crystal, set in a standard aluminum can, was top-loaded into the 6~T vertical field magnet with the field parallel to the 001 axis, or into the ILL-type Orange cryostat if external field is not necessary.
The elastic experiment has been performed also using ISSP-GPTAS, with a vertically focusing (horizontally flat) PG 002 monochromator; the spectrometer was operated in the double axis mode without the analyzer.
The incident neutron energy was selected as $E_{\rm i} = 14.79$~meV, calibrated using a standard Al$_2$O$_3$ powder sample.
Several collimation conditions were employed depending on necessity of $Q$ resolution, and two PG filters were inserted to eliminate the higher harmonic neutrons completely.

\section{Experimental results and discussion}
\begin{figure}
  \includegraphics[scale=0.32, angle=-90]{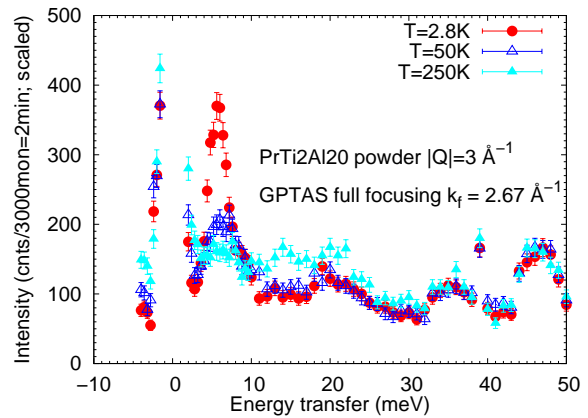}
  \caption{(color online) Inelastic spectra in a wide energy range up to 50~meV measured at the three temperatures $T = 2.8$, 50, and 250~K, and at the momentum transfer $Q = 3$~\AA$^{-1}$.
    We found that all the peaks above 10~meV ({\it i. e.} at 15, 20, 35, and 47~meV) are due to phonon scattering, judging from their temperature and $Q$ dependence.
  }\label{fig:wideEinelastic}
\end{figure}

\subsection{Inelastic scattering experiment on CEF excitations}

First, we measured the inelastic scattering spectra in wide temperature and energy ranges using a polycrystalline sample, aiming at determining the CEF level scheme of the Pr$^{3+}$ ions.
Figure~1 shows the neutron inelastic scattering spectra at the three temperatures $T = 2.8$, 50, and 250~K, and at the momentum transfer $Q = 3$~\AA$^{-1}$.
Strongly temperature dependent peak was observed around $\hbar \omega \simeq 6$~meV, whereas only weak, or negligible, temperature dependence was seen for the peaks above 17~meV.
From their $Q$-dependence (not shown), we concluded that the peaks above 17~meV are due to phonon contributions.
We also found that the intensity increase around $\hbar \omega \simeq 15$~meV at 250~K can be roughly reproduced by the Bose temperature factor ($[1-\exp(\hbar\omega/k_{\rm B}T)]^{-1}|_{T = 250~{\rm K}} \sim 2$), and thus this spectral weight is also due to the phonon scattering.
Therefore, we conclude that (observable) magnetic excitations exist only in the low energy region $\hbar \omega < 10$~meV.

Next, we measured the temperature dependence of the inelastic spectra in the low energy region.
Shown in Fig.~2 are the low-energy spectra in a wide temperature range observed at $Q = 1.5$~\AA$^{-1}$.
Before discussing the temperature dependence in detail, we first make a comment on the higher energy upturn commonly seen in all the spectra [except the one in Fig.~2(g), which will be described later].
This is due to the increase of background from contaminating main beam; at high energies $\hbar \omega > 12$~meV and at a relatively low $Q = 1.5$~\AA$^{-1}$, the scattering angle becomes quite low, and thus with the horizontally focusing analyzer the contamination from the main beam becomes serious, giving rise to this increase of the background level.
This background was separately estimated, and removed in the following fitting procedure.

Other than the upturn, there appear two inelastic peaks in the spectra.
These peaks were clearly observed at $T = 4.2$~K ($> T_{\rm c}$); an relatively sharp inelastic excitation peak was observed at $\hbar \omega \simeq 6$~meV, whereas an additional broad peak was observed as a hump around $\hbar \omega \simeq 9.5$~meV, as shown in Fig.~2(f).
The spectra at the elevated temperatures are shown in Fig.~2(a-e).
As the temperature is increased, the 6~meV and 9.5~meV peaks become weaker and broader.
Nonetheless, they are still observable at high temperatures as $T = 50$~K, indicating that they are robust excitations, originating from single-site effect.
Together with their $Q$ dependence (not shown), we conclude that they are CEF excitation peaks.

To check the change of the CEF excitation spectrum across the non-magnetic transition $T_{\rm c}$, the inelastic spectra were measured above and below $T_{\rm c}$ with higher statistical accuracy.
Shown in Fig.~2(g) are the two spectra at $T = 0.77$~K and 4~K.
For these scans, the supplemental radiation shield was placed along the main beam path, which completely suppresses the upturn of the background.
As is apparent in the figure, there is no detectable difference in the peak width as well as intensity for the two spectra within the present energy resolution and statistical accuracy.
As we will see in the next subsection, the CEF excitations have considerably large intrinsic width (of the order of a few meV) even at low temperatures.
It is, therefore, reasonable that a small change of energy levels due to the ordering at 2~K (corresponding to an energy scale of roughly 0.2~meV) cannot affect the peak profile significantly.

\begin{figure}
  \includegraphics[scale=0.32, angle=-90]{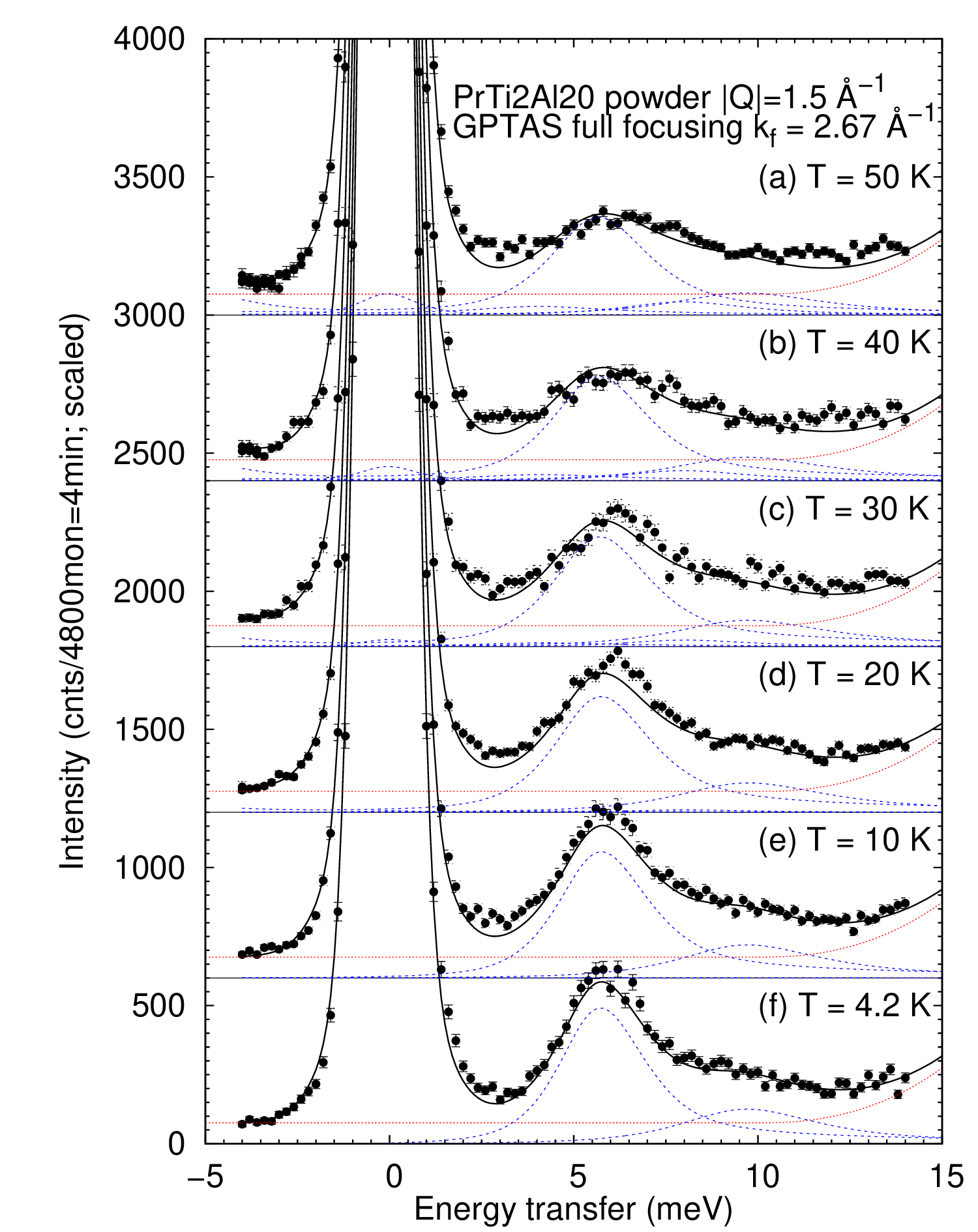}
  \includegraphics[scale=0.32, angle=-90]{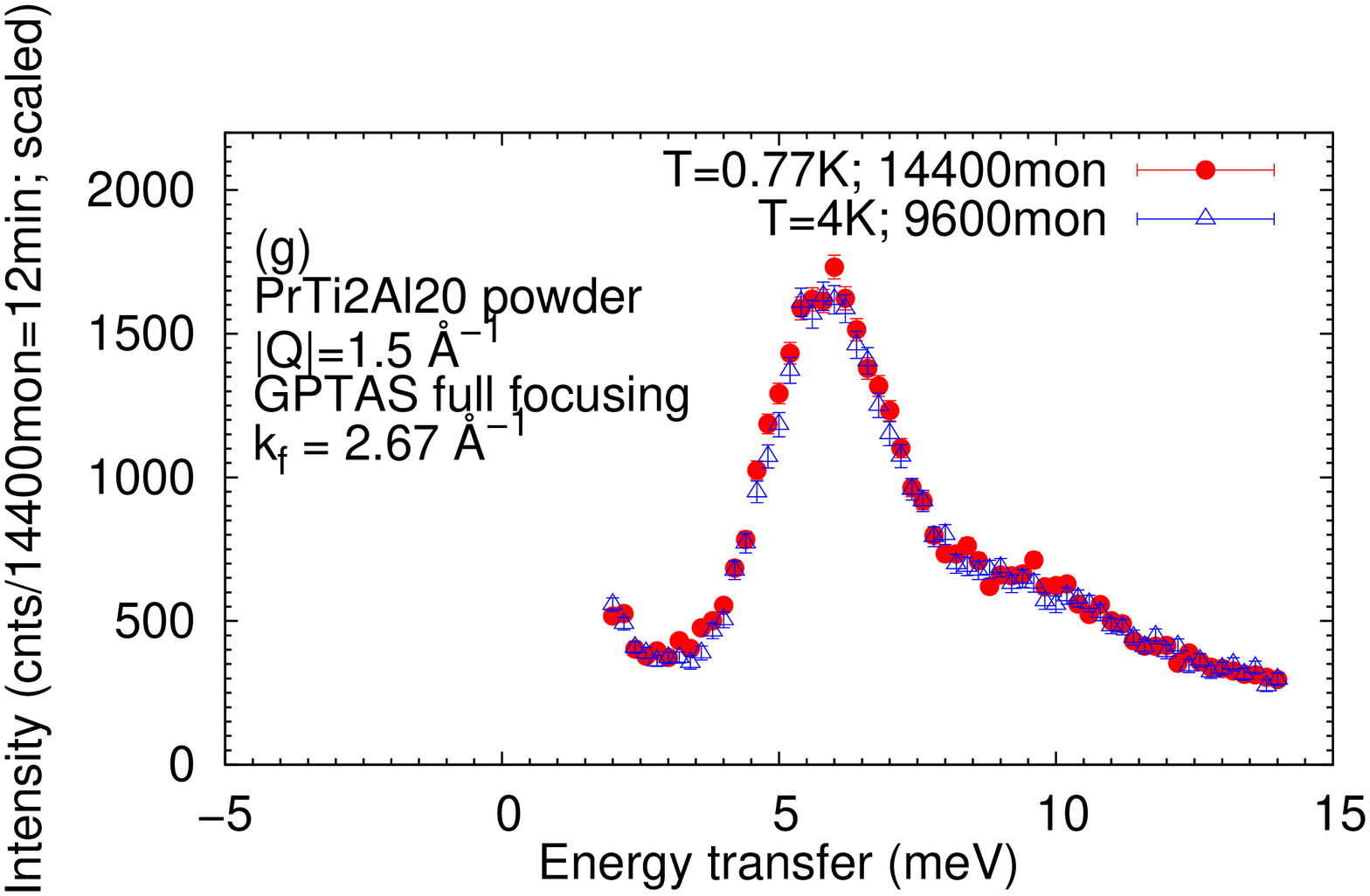}
  \caption{(color online) (a-f) Inelastic spectra at (a) $T = 50$~K, (b) $T = 40$~K, (c) $T = 30$~K, (d) $T = 20$~K, (e) $T = 10$~K, and (f) $T = 4.2$~K.
    The momentum transfer was fixed to $Q = 1.5$~\AA$^{-1}$.
    The solid lines stand for the total fitting results, whereas dotted lines represent each CEF peak, and dash-dotted line the background.
    The upturn of the background at higher energy ($\hbar \omega > 12$~meV) is due to the contamination from the main (direct) beam.
    See text for details.
    (g) Higher statistics data at $T = 0.77$~K ($< T_{\rm c}$) and $T = 4$~K ($> T_{\rm c}$).
    For these data, supplemental radiation shield was placed at low $2\theta$ to reduce the higher energy upturn of the background.
  }\label{fig:INSQ1.5}
\end{figure}

\subsection{Determination of CEF Hamiltonian parameters}
For the single Pr$^{3+}$ ion (the total angular momentum $J = 4$) under the point symmetry $\bar{4}3m$ ($T_{d}$), the CEF Hamiltonian may be written as:~\cite{LeaKR62}
\begin{equation}
{\cal H}_{\rm CEF} = W \left [ 
  x \frac{O_{40} + 5 O_{44}}{F_4} + (1-|x|) \frac{O_{60} - 21 O_{64}}{F_6}
  \right ],
\end{equation}
where $O_{40}, O_{44}, O_{60}$, and $O_{64}$ are the Stevens operator equivalents,~\cite{HutchingsMT64} and $F_4 = 60$ and $F_6 = 1260$ are the factors given in Ref.~[\onlinecite{LeaKR62}].
CEF removes the ninefold degeneracy of the $4f^2$-electron $J$-multiplet into four levels $|n\rangle$ ($n = 1,...,4$), corresponding to the irreducible representations $\Gamma_3, \Gamma_4, \Gamma_5$, and $\Gamma_1$ of $T_d$, respectively.
The transition strengths between these CEF splitting levels are then calculated as:
\begin{eqnarray}\label{eq:CEF}
b^{\alpha}_{nm} &=& \frac{2{\rm e}^{-E_n/k_{\rm B}T}}{Z} 
\frac{|\langle n |J^{\alpha}| m \rangle|^2}{E_m - E_n},\ \ \ (m \neq n)\nonumber\\
b^{\alpha}_{nn} &=& \frac{{\rm e}^{-E_n/k_{\rm B}T}}{Z} 
\frac{|\langle n |J^{\alpha}| n \rangle|^2}{k_{\rm B}T},\ \ \ \mbox{(otherwise)}
\end{eqnarray}
where $\alpha = x, y$, or $z$, and $k_{\rm B}$ and $Z$ are the Boltzmann constant and the partition function, respectively.
The scattering function from a powder sample may be given by a sum of spectral weights of the CEF transitions:
\begin{widetext}
\begin{equation}\label{eq:sqwWideE}
S(Q,\hbar \omega)_{\rm inel} = \frac{2}{3}\left [ \frac{1}{2} g_J f_{\rm mag}(Q) \right ]^2 \frac{N \hbar \omega}{1 - \exp(-\hbar \omega/k_{\rm B} T)}\sum_{nm\alpha}b^{\alpha}_{nm} P_{nm}(\hbar \omega; \hbar\omega_{nm},\Gamma_{nm}),
\end{equation}
where $N$, $f_{\rm mag}(Q)$, and $g_{\rm J}$ are the number, the magnetic form factor,~\cite{LisherEJ71} and the Lande $g$ factor of the Pr$^{3+}$ ions, respectively.
In the present analysis, we assume a pseudo-Voigt function as a profile function of the inelastic peaks, which is a reasonable approximation of an intrinsic Lorentzian-shaped excitation convoluted by a Gaussian-shaped instrumental resolution:~\cite{thompson87,wertheim74}
\begin{eqnarray}
P_{nm}(\hbar \omega; \hbar \omega_{nm}, \eta, \Gamma) &
= &\frac{1-\eta}{\Gamma}\sqrt{\frac{\ln 2}{\pi}} \exp\left[-4 \ln 2 \frac{(\hbar \omega - \hbar \omega_{nm})^2}{\Gamma^2}\right]
+ \frac{\eta}{\pi \Gamma} \frac{1}{(\Gamma/2)^2 + (\hbar \omega - \hbar \omega_{nm})^2}\\
&+& \frac{1-\eta}{\Gamma}\sqrt{\frac{\ln 2}{\pi}} \exp\left [-4 \ln 2 \frac{(\hbar\omega + \hbar\omega_{nm})^2}{\Gamma^2} \right ]
+ \frac{\eta}{\pi \Gamma} \frac{1}{(\Gamma/2)^2 + (\hbar\omega  + \hbar\omega_{nm})^2},
\end{eqnarray}
where, 
\begin{equation}
	\Gamma = (\Gamma_{\rm G}^5 + 2.69269 \Gamma_{\rm G}^4 \Gamma_{\rm L} + 2.42843 \Gamma_{\rm G}^3 \Gamma_{\rm L}^2 + 4.47163 \Gamma_{\rm G}^2 \Gamma_{\rm L}^3 + 0.07842 \Gamma_{\rm G} \Gamma_{\rm L}^4 + \Gamma_{\rm L}^5)^{1/5},
\end{equation}
\begin{equation}
  \eta = 1.36603 (\Gamma_{\rm L}/\Gamma) - 0.47719 (\Gamma_{\rm L}/\Gamma)^2 + 0.11116 (\Gamma_{\rm L}/\Gamma)^3.
\end{equation}
In the above equations, $\Gamma_{\rm L}$ is the full-width at half-maximum (FWHM) of the intrinsic Lorentzian-shaped spectral weight function of the CEF excitations.
For the width of the Gaussian-shaped instrumental resolution, $\Gamma_{\rm G}$, we assume energy dependent $\Gamma_{\rm G}(\hbar\omega_{nm}) = \Delta\hbar\omega_0(E_{\rm f} + \hbar\omega_{nm})/E_{\rm f}$ with the elastic width $\Delta\hbar\omega_0 = 1.2$~meV determined using the vanadium standard.
\end{widetext}

By performing least-square fitting to the observed spectra in a temperature range of $4.2 \leq T \leq 50$~K simultaneously, we obtained the optimum CEF parameters as: $x = 0.25(1)$ and $W = -1.53(3)$~meV.
In the fitting procedure, the Lorentzian widths $\Gamma_{\rm L}$ for the excitations between the ground state and excited states are set as adjustable, however $\Gamma_{\rm L}$ between the excited states are fixed to 3~meV, since they appear only in the spectra at high temperatures with weak intensity.
The fitting results for all the spectra with different temperatures are shown by the solid lines in Fig.~2(a-f), whereas the profiles of each CEF excitation peak and the background are given by the dotted and dash-dotted lines.
The reasonable coincidence between the calculated and observed spectra in a wide temperature range validates the obtained CEF parameters.
Temperature dependence of the peak widths between the ground state and the first excited state ($|1\rangle \rightarrow |2\rangle$) and that between the ground state and the second excited state ($|1\rangle \rightarrow |3\rangle$) is shown in Fig.~3.
The uncertainty of the width for the higher energy peak is considerably large, and thus we can hardly discuss the its temperature dependence.
On the other hand, the width for the lower energy peak exhibits clear decreasing behavior as temperature is lowered.
Interestingly, both the widths remain considerably large even at the lowest temperature.
This may suggest remaining dipole/quadrupole fluctuations due to the coupling to the conduction electrons.
However, it should be noted that such broad peaks may also originate from finite dispersion of the CEF excitations.
At the present moment we think this possibility is not likely; if this is the case, the peak should be narrower at higher temperatures, where the dispersion becomes weaker.
However, to unambiguously settle this issue, future single-crystal inelastic scattering study is necessary.

The resulting energy level scheme is illustrated in Fig.~4.
The ground state is the non-magnetic but quadrupolar (and octupolar) active $\Gamma_3$ doublet.
The first and second excited states are both magnetic triplets belonging to the $\Gamma_4$ (5.61~meV) and $\Gamma_5$ (9.30~meV) irreducible representations.
The highest-energy excited state is consequently the non-magnetic $\Gamma_1$ singlet state at 13.47~meV, which cannot be be seen in a neutron scattering spectrum.
The list of determined energies levels and corresponding wave functions is given in Table.~I.

\begin{figure}
  \includegraphics[scale=0.32, angle=-90]{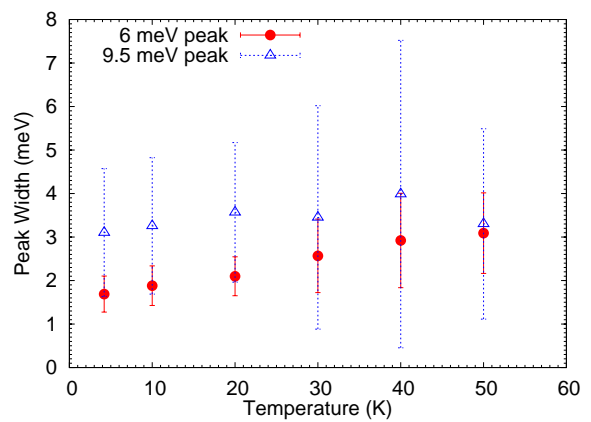}
  \caption{(color online) Temperature dependence of the width parameters, $\Gamma_{\rm L}$, for the 6~meV and 9.5~meV peaks.
    Note that the uncertainty is considerably large for the higher-energy-peak width, prohibiting us to discuss its temperature dependence.
  }\label{fig:gammaTdep}
\end{figure}

\begin{figure}
  \includegraphics[scale=0.38, angle=-90]{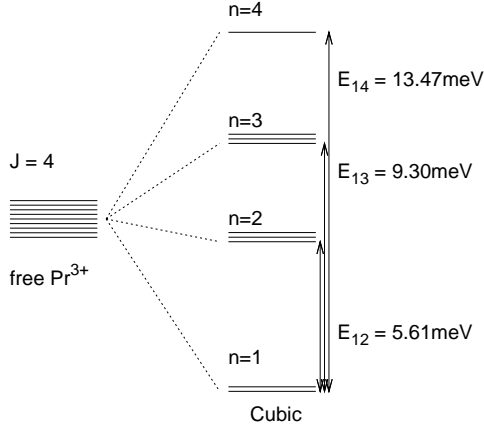}
  \caption{Determined CEF level scheme for PrTi$_2$Al$_{20}$.
  }\label{fig:CEFscheme}
\end{figure}

\begin{table}  
\caption{List of determined energy levels ($E_{1n}$) and corresponding wave functions for the Pr$^{3+}$ ions in PrTi$_2$Al$_{20}$.
  $E_{nm}$ is defined as $E_{nm} = E_m - E_n$, whereas $n$ $(m)$ stands for the numbers to specify energy levels, used in Fig.~4.
  }
  \begin{tabular}{cccc}
    \hline
    $n$ & $E_{1n}$ (meV) & Irrep & Wave functions\\
    \hline
    4 & 13.47 & $\Gamma_1$ & $\frac{1}{2}\sqrt{\frac{5}{6}} |4\rangle + 
    \frac{1}{2}\sqrt{\frac{7}{3}}|0\rangle +
    \frac{1}{2}\sqrt{\frac{5}{6}} |-4\rangle $\\

    3 & 9.3 & $\Gamma_{5\pm}^{(1)}$ & $\frac{1}{2}\sqrt{\frac{7}{2}} |\pm 3\rangle - 
    \frac{1}{2}\sqrt{\frac{1}{2}}|\mp 1\rangle$\\

      &     & $\Gamma_{5}^{(2)}$ & $\sqrt{\frac{1}{2}} |2\rangle - \sqrt{\frac{1}{2}}|-2\rangle$\\

    2 & 5.61 & $\Gamma_{4\pm}^{(1)}$ & $\frac{1}{2}\sqrt{\frac{1}{2}} |\mp 3\rangle + 
    \frac{1}{2}\sqrt{\frac{7}{2}}|\pm 1\rangle$\\

      &     & $\Gamma_{4}^{(2)}$ & $\sqrt{\frac{1}{2}} |4\rangle - \sqrt{\frac{1}{2}}|-4\rangle$\\

    1 & 0 & $\Gamma_{3}^{(1)}$ & $\frac{1}{2}\sqrt{\frac{7}{6}} |4\rangle - 
    \frac{1}{2}\sqrt{\frac{5}{3}}|0\rangle + \frac{1}{2}\sqrt{\frac{7}{6}} |-4\rangle $\\
      &     & $\Gamma_{3}^{(2)}$ & $\sqrt{\frac{1}{2}} |2\rangle + \sqrt{\frac{1}{2}}|-2\rangle$\\
    \hline
  \end{tabular}
\end{table}

\subsection{Order parameter}

\begin{figure}
  \includegraphics[scale=0.165, angle=-90]{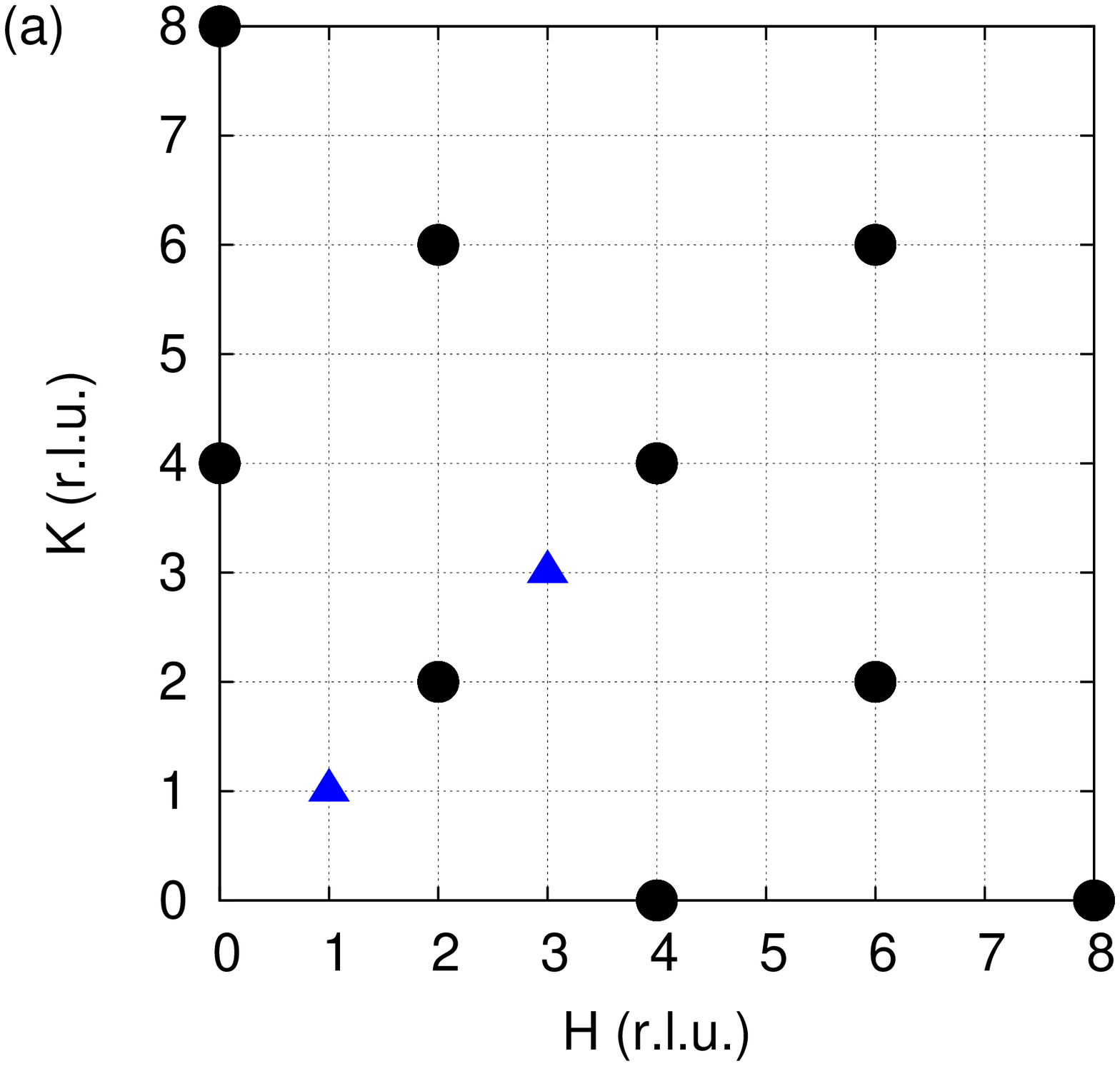}
  \includegraphics[scale=0.165, angle=-90]{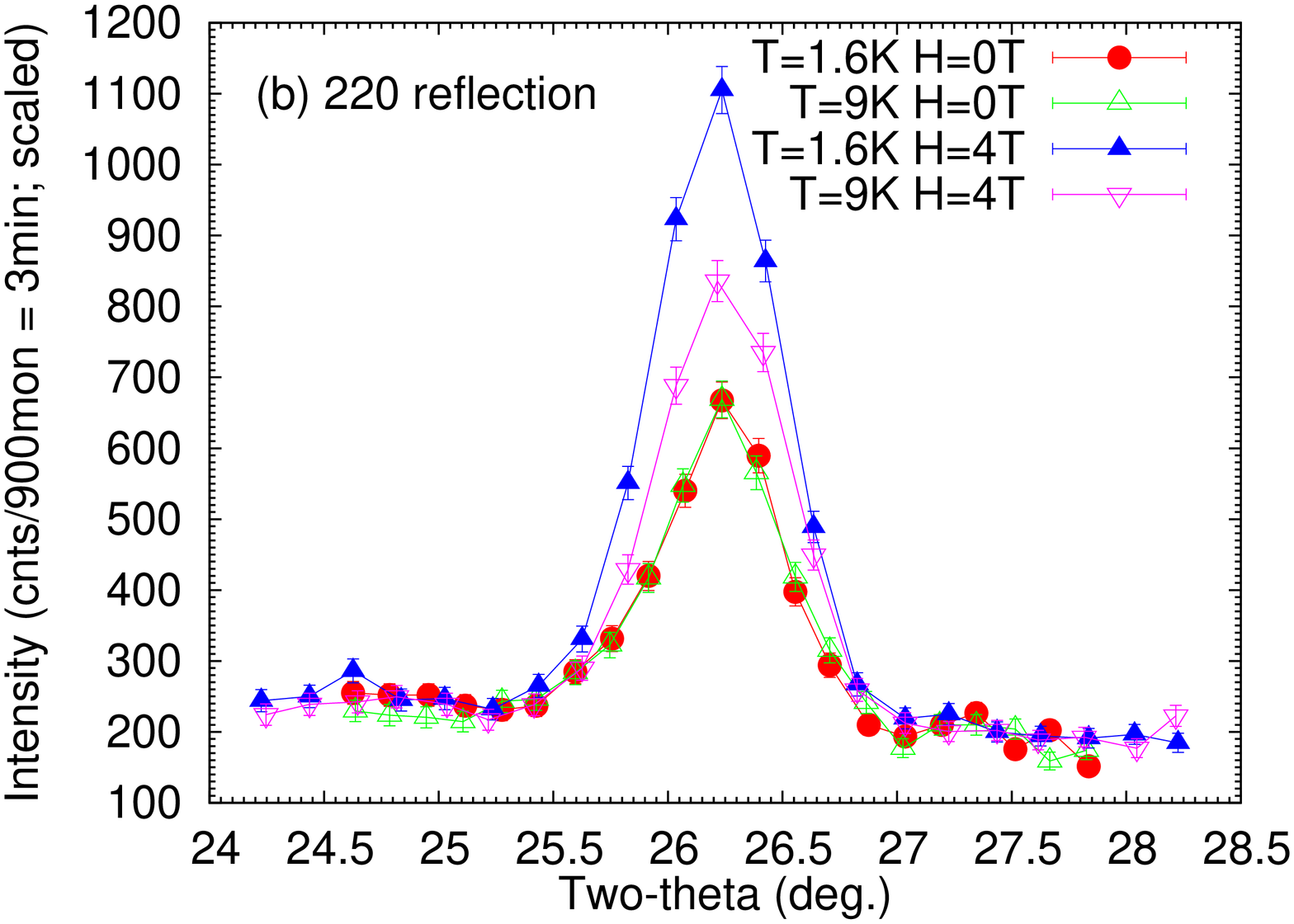}
  \includegraphics[scale=0.165, angle=-90]{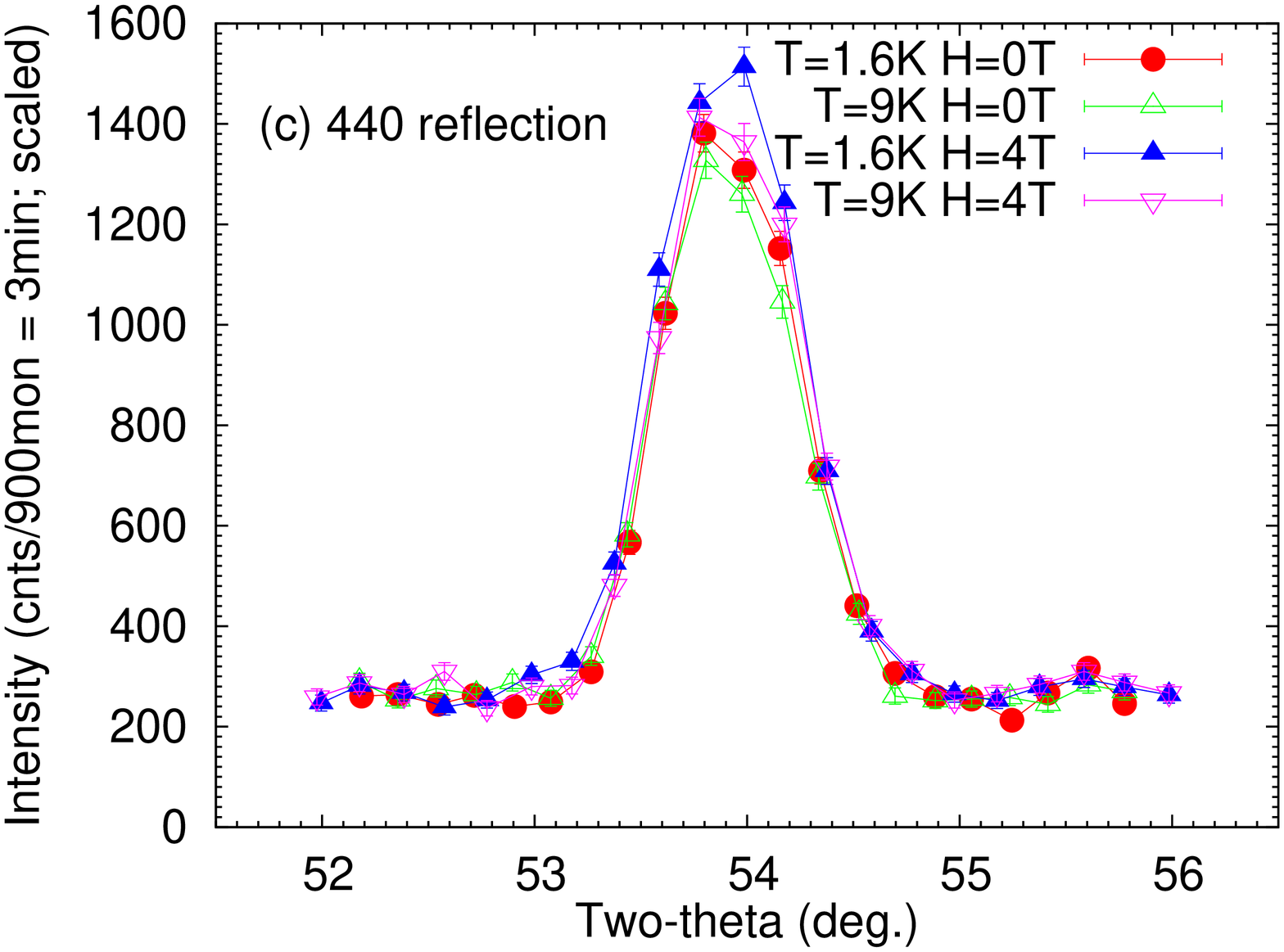}
  \includegraphics[scale=0.165, angle=-90]{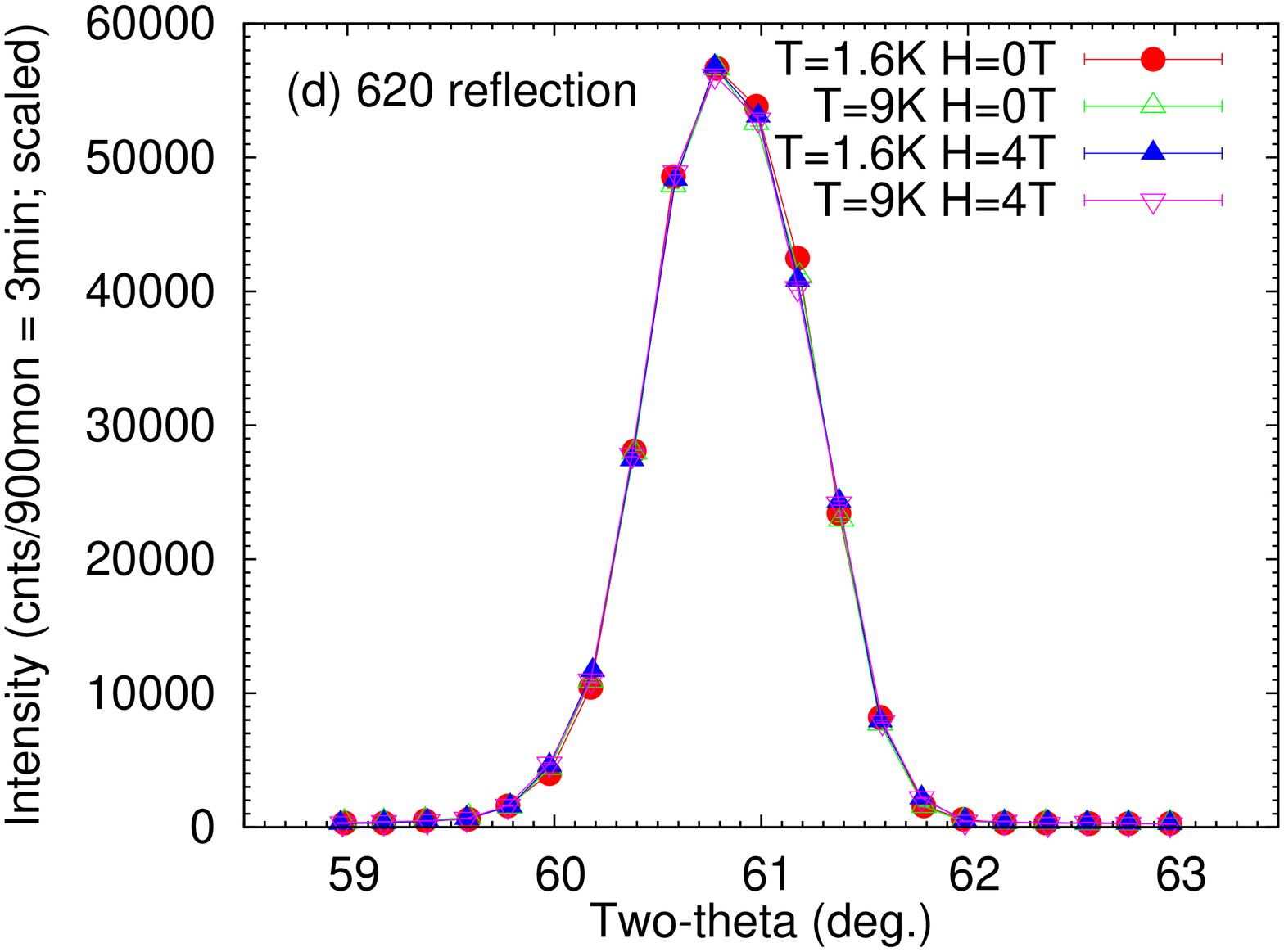}
  \includegraphics[scale=0.165, angle=-90]{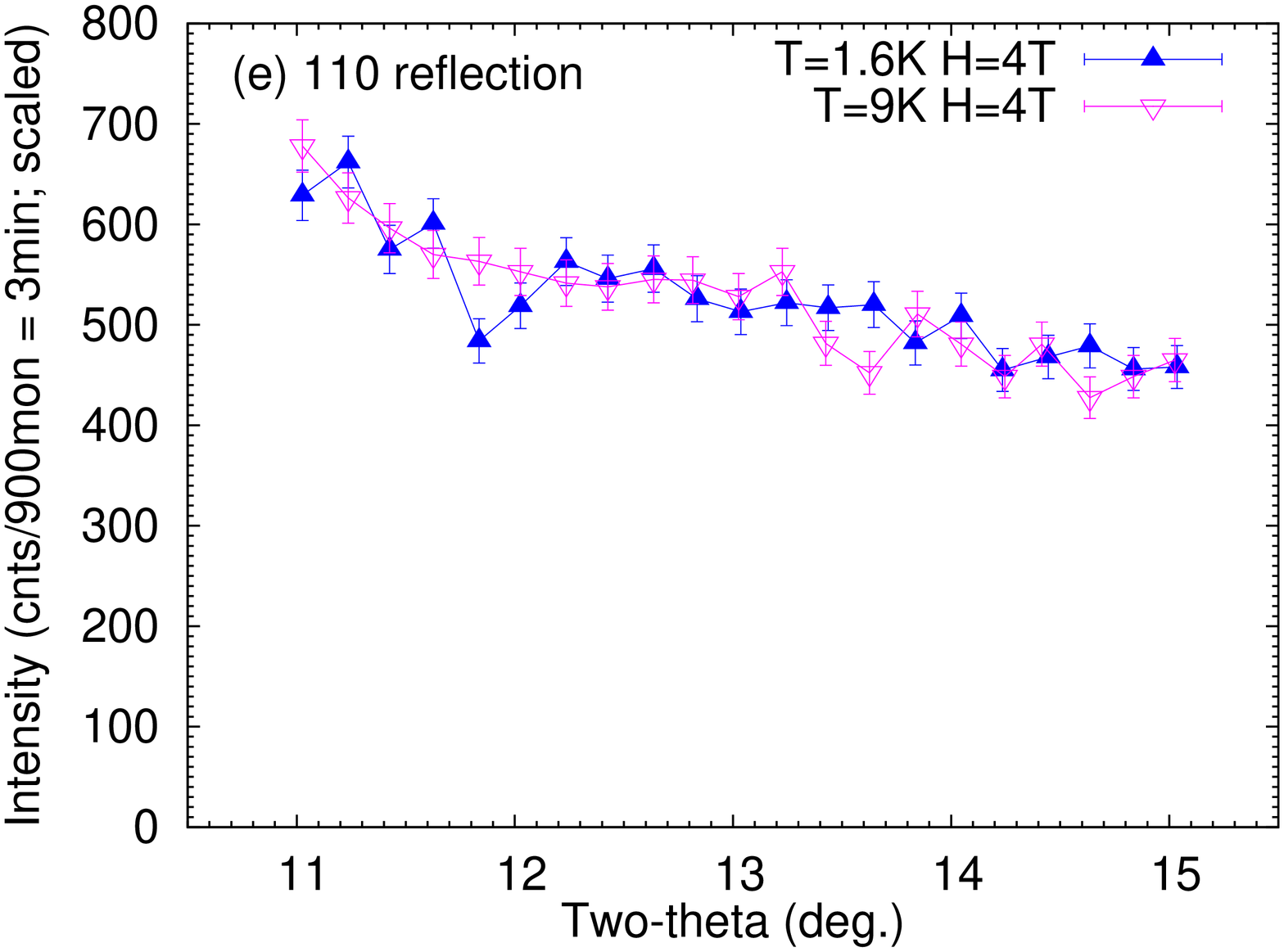}
  \includegraphics[scale=0.165, angle=-90]{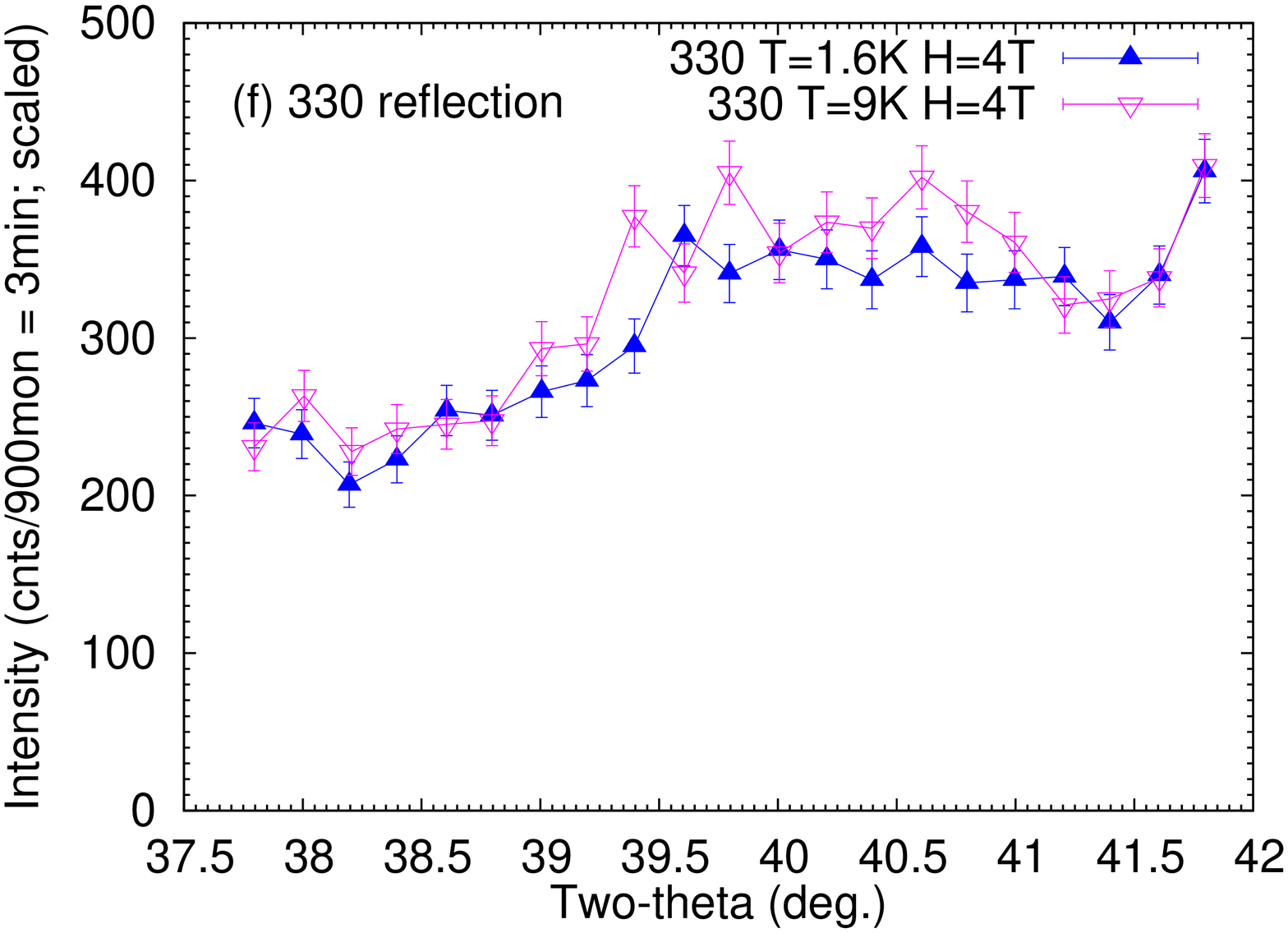}
  \caption{(color online) (a) $Q$ positions where the reflection intensity was collected in the present study.
    (b-f) $\theta-2\theta$ scans at representative $Q$ positions, 220, 440, 620, 110 and 330.
    For the 220, 440, and 620 reflections, both the zero-field and 4~T data were collected at both $T = 9~{\rm K} > T_{\rm c}$ and $T = 1.6~{\rm K} < T_{\rm c}$, whereas for the 110 and 330 reflections only $H=4$~T data were measured.
  }\label{fig:theta2thetascans}
\end{figure}

It has been confirmed in our neutron inelastic experiment that the ground state is non-magnetic $\Gamma_3$ doublet in PrTi$_2$Al$_{20}$, consisting of the two wave functions, $\Gamma_3^{(1)}$ and $\Gamma_3^{(2)}$, as listed in Table I.
For the $\Gamma_3$ doublet, two components of quadrupolr moments, $O_{20} = \frac{1}{2}(3 J_z^2 - J^2)$ and $O_{22} = \frac{\sqrt{3}}{2}(J_x^2 - J_y^2)$, are allowed to be finite, as well as one higher-order octupolar moment, $T_{xyz} = \frac{\sqrt{15}}{6} \overline{J_x J_y J_z}$.
(The bar stands for possible permutation of the three operators.)
Hence, one of them may be the order parameter responsible for the mysterious non-magnetic transition observed around 2~K in the macroscopic measurements.
The quadrupolar and octupolar order parameters cannot be directly measured using neutron scattering because of the absence of the significant coupling between the neutron and the multipole moments.
Nevertheless, magnetic (dipole) moment, neutron-observable, can be induced by mixing the excited magnetic states into the non-magnetic ground state under finite external magnetic field.
The mixing, and consequently the size and direction of the induced moment, depend on the symmetry of the ground- and excited-state wave functions.
Therefore, one may distinguish the wave-function symmetry, and accordingly the symmetry of the order parameter, using the neutron scattering under external field.~\cite{EffantinJM85}

Much specifically, for the finite order parameter $\langle O_{20} \rangle \neq 0$, the degenerated $\Gamma_3$ wave functions split into two as $\Gamma_3^{(1)}$ and $\Gamma_3^{(2)}$.
On the other hand for $\langle O_{22} \rangle \neq 0$, linear combinations of the $\Gamma_3$ wave functions, $\frac{1}{\sqrt{2}}(\Gamma_3^{(1)} + \Gamma_3^{(2)})$ and $\frac{1}{\sqrt{2}}(\Gamma_3^{(1)} - \Gamma_3^{(2)})$ will be the energy-split eigenfunctions.
For $\langle T_{xyz} \rangle \neq 0$, $\Gamma_3$ will split into $\frac{1}{\sqrt{2}}({\rm i} \Gamma_3^{(1)} + \Gamma_3^{(2)})$ and $\frac{1}{\sqrt{2}}(-{\rm i} \Gamma_3^{(1)} + \Gamma_3^{(2)})$.
Under the magnetic field along 001 ({\it i.e.} $H \parallel z$), the $\langle O_{20}\rangle$ order parameter will give rise to extra (additional) induced dipole moment along $z$, $<J_z>$, whereas for the rest, no coupling between the multipolar order and induced dipole moment is expected.~\cite{ShiinaR97}
Therefore, we may distinguish $\langle O_{20} \rangle$ by observing appearance of extra induced moment under the external field along $z$ in the ordered phase.

With the above expectation in mind, we have performed neutron diffraction experiment under magnetic field along $z$.
A number of $Q$ positions in the $hk0$ plane were investigated using the $\theta-2\theta$ scans, with or without the external magnetic field, and at $T = 1.6$~K ($< T_{\rm c}$) as well as 9~K ($> T_{\rm c}$).
The investigated $Q$ positions are shown in Fig.~5(a); closed circles stand for the positions where the nuclear Bragg reflections are allowed, whereas triangles denote nuclear forbidden positions.

Representative results of the $\theta-2\theta$ scans are shown in Fig.~5(b-f).
At the 220 position shown in Fig.~5(b), results at 1.6~K and 9~K under zero field perfectly coincides with each other, confirming the absence of the ferromagnetic dipolar ordering.
By applying $H = 4$~T along the 001 direction at $T = 9$~K, the scattering intensity increases weakly, indicating that the ferromagnetically aligned dipole moments are induced by the external magnetic field.
As the temperature is lowered to $T = 1.6$~K, significant increase of the scattering intensity was observed, suggesting an appearance of the extra induced moment.
The temperature dependence of the scattering intensity will be discussed later in detail.

Such an increasing behavior of the scattering intensity below $T_{\rm c}$ under the finite field is also detected for the 440 reflection, shown in Fig.~5(c).
Since the nuclear Bragg reflection intensity, which is supposed to be temperature independent in this low temperature range, is relatively strong compared to the 220 reflection, the increase of the scattering intensity at $T = 1.6$~K (and at $H = 4$~T) is less prominent for 440.
Nevertheless, the increase is roughly 200 counts/3~minutes, which is indeed a similar value as we observed for the 220 reflection.
Such increase of the reflection intensity was not observed at other nuclear allowed positions, as exemplified by the 620 reflection shown in Fig.~5(d).
However, this is simply due to their much stronger nuclear intensity compared to the magnetic signal; we cannot obtain necessary statistical accuracy to detect magnetic signal for these reflections within a reasonable experimental time.

The nuclear forbidden positions, such as the 110 and 330 reflections, were then checked in a similar manner.
Resulting $\theta-2\theta$ scans are shown in Fig.~5(e) and 5(f).
As can be seen in these figures, no increase of the scattering intensity was detected as temperature is decreased to $T = 1.6$~K even under the finite external field $H = 4$~T.
From these observations, we conclude that the increase of the scattering intensity under finite external field below $T_{\rm c}$ can be observed only on top of the allowed nuclear Bragg reflections, and therefore the ordering in PrTi$_2$Al$_{20}$ cannot be staggered one, but is ferro-type ordering. 

\begin{figure}
  \includegraphics[scale=0.35, angle=-90]{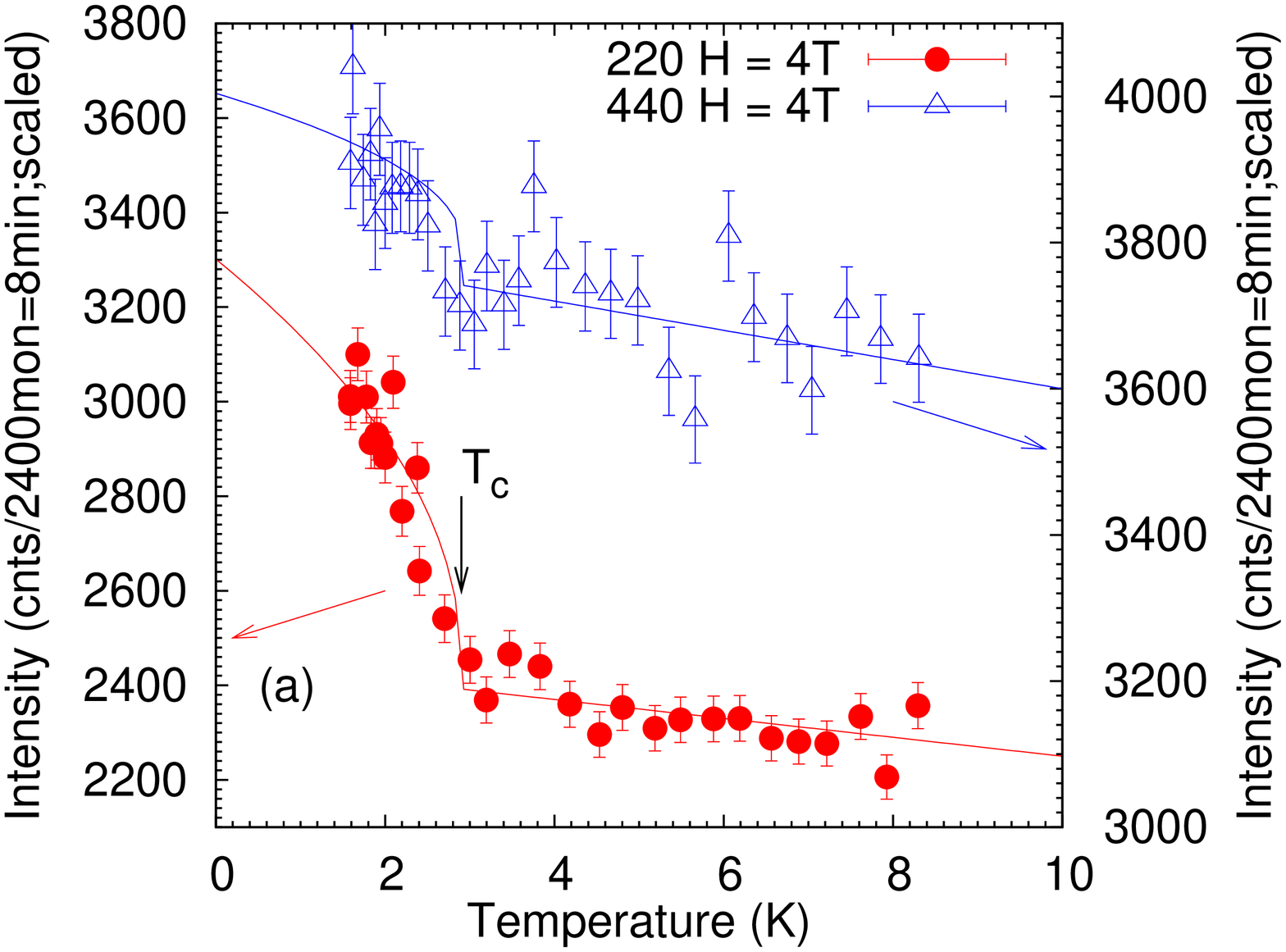}
  \includegraphics[scale=0.35, angle=-90]{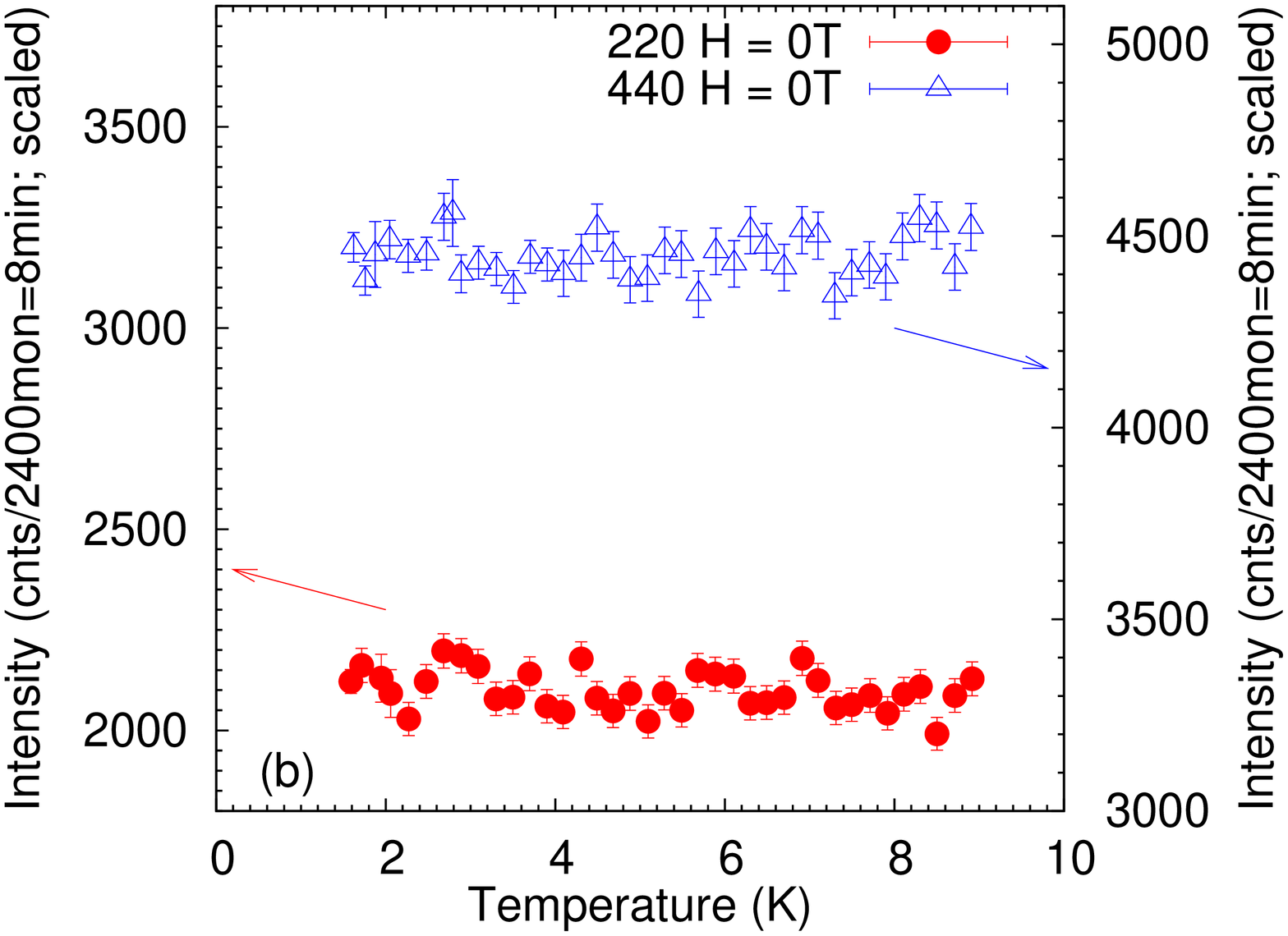}
  \caption{(color online) (a) Temperature dependence of the 220 (closed circles) and 440 (open triangles) reflection intensity under the external magnetic field $H = 4$~T.
    Solid lines are guides to the eyes.
    (b) Temperature dependence of the 220 (closed circles) and 440 (open triangles) reflection intensity under the zero external magnetic field.
    Absolute counting number inconsistency between the $H = 4$~T and $H = 0$~T results is due to the difference in the used cryostats; for the $H = 4$~T experiment, we used the vertical field magnet which has thicker radiation shields with several dark angles, whereas we used the simpler Orange cryostat for the zero-field experiment.
  }\label{fig:orderparameters}
\end{figure}

\begin{figure}
  \includegraphics[scale=0.32, angle=-90]{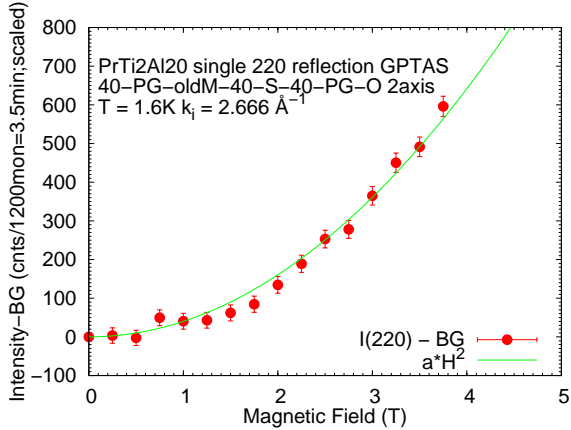}
  \caption{(color online) Magnetic field dependence of the 200 reflection intensity measured at the base temperature $T = 1.6$~K.
    The nuclear scattering intensity, estimated at the zero field, is subtracted as a nuclear scattering background.
    Solid line stands for a result of fitting to the $H^2$ dependence.
  }\label{fig:Hdep}
\end{figure}

To see if the increase of the scattering intensity under finite external field is indeed a signature of a phase transition, we have checked the temperature dependence of the reflection intensity, which is, in the present case, the square of the induced moment.
The temperature dependence of the 220 and 440 reflection intensity under $H = 4$~T is shown in Fig.~6(a).
Drastic increase can be see at $T_{\rm c} \simeq 2.8$~K, indicating that cooperative ordering of the induced magnetic moments take place at this temperature.
Shown in Fig.~6(b) are the corresponding temperature dependence of the reflection intensity measured under zero external field.
As is clearly seen, there appears no critical increase of the reflection intensity, indicating that there is no ordering of the dipole moments under zero field.

External field dependence of the 220 reflection intensity is shown in Fig.~7.
Apparently, the reflection intensity shows continuous $H^2$ dependence.
This certifies that the phase at $H = 4$~T and $T < T_{\rm c}$~K is continuous to that at zero field.
In other words, the critical behavior observed under $H = 4$~T is not due to the field-induced ordering of dipole moments, but due to the spontaneous ordering of quadrupole (or octupole) moments, which was observed as the non-magnetic anomaly in the macroscopic measurements in zero field.
By normalizing the field-induced intensity of the 220 reflection at $H = 4$~T in the ordered phase ($T = 1.6$~K) using the purely nuclear reflection intensity obtained at $T = 9$~K and $H = 0$~T, we estimated the size of the induced moment as 0.41(3)~$\mu_{\rm B}$.
This is in a good agreement with those obtained in the similar (but much complicated incommensurately ordered) intermetallic compound PrPb$_3$.~\cite{OnimaruT05}

The critical increase of the field-induced dipole moment was clearly observed for $H \parallel z$ below $T_{\rm c}$ as shown in Fig.~6(a). 
As mentioned earlier, this increase can be expected only in an ordered phase with finite $\langle O_{20} \rangle$ quadrupolar order parameter, but not for $\langle O_{22} \rangle$ nor $\langle T_{xyz} \rangle$.
The observed $H^2$ dependence of the squared induced-moment (Fig.~7) is also what is expected for the $\langle O_{20} \rangle$ ordered phase.
We, hence, conclude that the ferro-quadrupolar ordered phase is established below $T_{\rm c}$ with the finite order parameter $\langle O_{20} \rangle$.

\section{Summary}
We have performed both the polycrystalline-inelastic and single-crystal-diffraction experiments on PrTi$_2$Al$_{20}$.
Two CEF peaks were observed in the inelastic experiment, which were assigned to the $\Gamma_4$ and $\Gamma_5$ levels of the Pr$^{3+}$ ions.
The ground state for $T > T_{\rm c}$ is confirmed to be the non-magnetic doubly degenerated $\Gamma_3$ state.
In the diffraction experiment, the field-induced dipole moment shows critical increase at $T_{\rm c}$, appearing on the top of the nuclear allowed reflections.
From those results, we conclude that the ferro-quadrupolar order with the finite order parameter $\langle O_{20} \rangle$ is established in PrTi$_2$Al$_{20}$.

\section*{Acknowledgments}
The present authors thank Prof. T. Onimaru for valuable discussions.
This work is partially supported by Grant-in-Aids for Scientific Research (No. 21684019 and No. 20560612) from JSPS, and by Grant-in-Aid for Scientific Research on Innovative Areas ``Heavy Electrons'' from MEXT, Japan.


\end{document}